\documentclass[aps,pra,twocolumn,superscriptaddress,showpacs,preprintnumbers,amsmath,amssymb]{revtex4-1}

\usepackage{graphicx}
\usepackage{dcolumn}
\usepackage{bm}
\usepackage{enumerate}
\usepackage{mathrsfs}
\usepackage{amsmath}

\usepackage{epstopdf}
\usepackage[titletoc]{appendix}
\usepackage[colorlinks=true,breaklinks=true,linkcolor=blue,citecolor=blue,urlcolor=blue]{hyperref}
\usepackage{color}
\usepackage{array,mathtools,amssymb,booktabs}
\newcolumntype{C}{>{$}c<{$}}
\AtBeginDocument{
\heavyrulewidth=.08em
\lightrulewidth=.05em
\cmidrulewidth=.03em
\belowrulesep=.65ex
\belowbottomsep=0pt
\aboverulesep=.4ex
\abovetopsep=0pt
\cmidrulesep=\doublerulesep
\cmidrulekern=.5em
\defaultaddspace=.5em
}

\begin{document}

\bibliographystyle{apsrev} 

 \def\rtx@apspra{\class@info{APS journal PRA selected}}

\title{Microgravity-assisted off-axis spin vortex in a $^{87}$Rb dipolar spinor Bose-Einstein condensate}

\author{Hui Tang}
\affiliation{Shenzhen Institute for Quantum Science and Engineering, Southern University of Science and Technology, Shenzhen, Guangdong 518055, China}

\author{Wenxian Zhang}
\email[Corresponding email: ]{wxzhang@hznu.edu.cn}
\affiliation{School of Physics, Hangzhou Normal University, Hangzhou, Zhejiang 311121, China}

\date{\today}

\begin{abstract}
The generation of the ground state of a spin vortex in a $^{87}$Rb Bose-Einstein condensate with the assistance of an optical plug has been studied. However, gravity is everywhere, and this potential linear dependence on the spatial position will destroy the axisymmetric structure of the system with the optical plug. In this case, the question of whether the spin vortex ground state still exists remains unresolved. The present study aims to explore the impact of microgravity on the formation of the spin vortex state with the assistance of an optical plug. To this end, a simple model has been employed to provide a comprehensive understanding of the phenomenon. The Gross-Pitaevskii equations are solved by setting the optical plug intensity, adjusting the optical plug width, and adjusting the microgravity strength. This process results in the phase diagram for the single-mode state and spin vortex state. Under microgravity situations, we observe an off-axis structure of the spin vortex state. Our calculations offer a reliable approach to generating spin vortex states in a microgravity environment.
\end{abstract}

\maketitle

\section{ Introduction}

Gravity is an important element that is noteworthy in the field of Bose-Einstein condensate(BEC). Gravity influences the atoms' cooling limit in experiments generating a BEC~\cite{KETTERLE1996181, Chen_2020, Guery-Odelin:98, leanhardt2003cooling, 2018nature, Corgier_2018}. The efficiency of evaporative cooling is directly impacted by gravity's effect, as it cannot be completely disregarded from the beginning of cooling~\cite{KETTERLE1996181, PhysRevA.73.043410}. A gradient magnetic field can be employed to offset the effects of gravity in single-component BEC experiments~\cite{leanhardt2003cooling}. However, a gradient magnetic field, on the other hand, may cause the internal spin components of BEC to split apart or even escape from the trap. Therefore, the BEC experiment urgently requires a microgravity environment.

In 2010, Rasel used the microgravity environment generated by free fall to cool a $^{87}$Rb BEC to the order of $10^{-9}$ K. Then, they first achieved a BEC cooling experiment below $1n$K~\cite{2018nature, Corgier_2018}. Now, cold atom experiments have already been carried out in microgravity environments in outer space. Due the microgravity environment, the bubble shape of BEC have been achived~\cite{carollo2022observation}. Additionally, the application of cold atoms in atomic clocks~\cite{SALOMON20011313,WOS:000607749700001,NewOa2023}  and atomic interferometers~\cite{PhysRevLett.110.093602} has been promoted by microgravity condition. With help of optical plug, one can generate a spin vortex ground state in a disc shape dipolar $^{87}$Rb BEC~\cite{PhysRevA.105.063324}. While the effect of the gravity have not been considered. Actually, the gravity don't change the shape of BEC under a harmonic trap. However, in order generation a spin vortex ground state, the optical plug is nececcery. 

In the presence of Earth's gravity and optical plug, the axial symmetry of the BEC breaks down along the z-axis (the direction of the highest frequency). This prevents the spin density from being uniform in the direction perpendicular to the density gradient. This spin density configuration corresponds to the minimum magnetic dipole-diplole interaction energy and can exist in an axially symmetric BEC. This spin structure's ground state can be achieved by adjusting the intensity and width of the optical plug~\cite{PhysRevA.105.063324}. Let us review the formation mechanism of this ground state with an optical plug. Without an optical plug, forming spin vortex states leads to a topologically non-trivial phase structure of the spin components $\pm1$, which causes these components to have zero density near the vortex core~\cite{PhysRevLett.97.020401, PhysRevLett.97.130404, sadler2006spontaneous}. This means that the $\vert 0\rangle$ component occupies the vortex core. This type of spin vortex state is called a polar core vortex(PCV). This phase separation prevents the three components from maintaining a density distribution of $n_1:n_0:n_-1=1:2:1$, meaning that the spin cannot be maximally polarized~\cite{PhysRevA.66.011601, PhysRevA.94.063615}. Consequently, the spin-spin contact energy ($E_2$) cannot simultaneously reach its minimum with the intrinsic dipole interaction energy ($E_{dd}'=0$)~\cite{PhysRevA.105.063324}. With optical plug, the density at the core centre is zero, so there is no competition between the $E_{dd}'$ and $E_2$. However, if a microgravity exists, the structure where $E_{dd}' = 0$, cannot be achived. We will now consider how the formation of spin vortex states is influenced by a microgravity environment. 

In this paper, we consider a $^{87}$Rb dipolar spinor BEC in a highly pancake optical trap~\cite{PhysRevLett.75.3969, PhysRevLett.104.160401, fu2003effect}, with microgavrity enviroment and an additional optical plug. To found the effect of microgravity, we adjust the width of optical plug and microgravtiy to find a phase transition in $^{87}$Rb dipolar condensate. The intensity of optical plug is setting to be a constant here. We utilize a simple model to investigate the effect of microgravity, which is confirmed in numerical computations by resolving Gross-Pitaevskii equations(GPEs) of a $^{87}$Rb dipolar spinor condensate with a disc shape. Numerical calculations demonstrate that the microgravity strength causes a phase transition from the spin vortex to the single-mode state(SMA)~\cite{PhysRevLett.93.040403}. Our findings offer a trustworthy method for researching phase transitons caused by microgravity in a $^{87}$Rb dipolar spinor condensate.

The paper is organized as follows. We intruduce the nonlocal system under the microgravity for the $^{87}$Rb dipolar spinor BEC in Sec.~\ref{sec:nonlocal}. The effect of microgravity is studied in a simple model, by calculating the energy difference of two local minimum states in Sec.~\ref{sec:osv}.  Then, the numerical results are presented in Sec.~\ref{sec:num}, which confirm the phase transition induced by microgravity. Conclusions are given  in Sec.~\ref{sec:conc}.


\section{Nonlocal dipolar model}
\label{sec:nonlocal}

The Hamiltonian of a $^{87}$Rb dipolar spin-1 BEC can be described as follows~\cite{PhysRevA.82.043627, PhysRevLett.97.130404, PhysRevA.73.023602, lahaye2009physics, PhysRevA.105.063324},
\begin{equation}
\begin{aligned}
\hat{H}=&\int d\mathbf{r} \left[\hat{\psi}^{\dag}_{\alpha}(\mathbf{r})\left(-\frac{\hbar^2\nabla^2}{2M}+V(\mathbf{r})\right)\hat{\psi}_{\alpha}(\mathbf{r})\right.\\
&+\frac{c_0}{2}\hat{\psi}^{\dag}_{\alpha}(\mathbf{r})\hat{\psi}^{\dag}_{\beta}(\mathbf{r})\hat{\psi}_{\beta}(\mathbf{r})\hat{\psi}_{\alpha}(\mathbf{r})\\
&+\left.\frac{c_2}{2}\hat{\psi}^{\dag}_{\alpha}(\mathbf{r})\hat{\psi}^{\dag}_{\alpha^{\prime}}(\mathbf{r})\mathbf{\textit{\textbf{F}}}_{\alpha\beta}\cdot\mathbf{\textit{\textbf{F}}}_{\alpha^{\prime}\beta^{\prime}}\hat{\psi}_{\beta}(\mathbf{r})\hat{\psi}_{\beta^{\prime}}(\mathbf{r})\right] \\
&+\hat{H}_{dd},
\label{eq:total}
\end{aligned}
\end{equation}
where $\hat{\psi}_{\alpha}$ $(\alpha=0, \pm1)$ represents the filed operator, $M$ the mass of a $^{87}$Rb atom, and $\boldsymbol{F}=(F_x,F_y,F_z)$ the spin-1 matrix, its three components are
\begin{equation}
	\begin{aligned}
	&F_x=	\frac{1}{\sqrt{2}}\begin{pmatrix}
	0 & 1 & 0\\
	1 & 0 & 1\\
	0 & 1 & 0
\end{pmatrix},
F_y=	\frac{i}{\sqrt{2}}\begin{pmatrix}
	0 & -1 & 0\\
	1 & 0 & -1\\
	0 & 1 & 0
\end{pmatrix},\\
&F_z=	\begin{pmatrix}
	1 & 0 & 0\\
	0 & 0 & 0\\
	0 & 0 & -1
\end{pmatrix},
	\end{aligned}
\end{equation}
local interaction parameters $c_0=4\pi\hbar^{2}(a_0+2a_2)/(3M)$, $c_2=4\pi\hbar^{2}(a_2-a_0)/(3M)$ represent the strength of spin-independent and spin-spin contact interaction, respectively. The scattering length of the total spin $s$ is represented as $a_s (s=0,2)$. The $V(\boldsymbol{r})$ is an external trap composed of a harmonic trap, an optical plug and microgravity with acceleration $a$.
\begin{equation}
\setlength\abovedisplayskip{3pt}
\setlength\belowdisplayskip{3pt}
V(\mathbf{r}) = \frac{1}{2}M\omega_0^{2}(x^2+y^2+\lambda^2z^2) +U_{0}e^{-(x^2+y^2)/\sigma^2}+Max,
\end{equation}
where $U_0$ and $\sigma$ are two parameters describing the barrier height and width of the Gaussian optical plug~\cite{PhysRevLett.75.3969, PhysRevLett.104.160401, fu2003effect}, respectively, and $\omega_0$ is the harmonic trap angular frequency in the $x$-$y$ plane and $\lambda$ the trap aspect ratio. The microgravity is along the x axis.

The last term of Eq. \ref{eq:total} describe magnetic dipole-dipole interaction(MDDI), 
\begin{equation}
\setlength\abovedisplayskip{3pt}
\setlength\belowdisplayskip{3pt}
\begin{split}
\hat{H}_{dd}=&\frac{c_{dd}}{2}\int d\mathbf{r}\int d\mathbf{r^{\prime}}\frac{1}{|\mathbf{r-r^{\prime}}|^{3}}\\
\times &[\hat{\psi}_{\alpha}^\dag(\mathbf{r})\hat{\psi}_{\alpha^{\prime}}^\dag(\mathbf{r})\mathbf{\textit{\textbf{F}}}_{\alpha\beta}\cdot\mathbf{\textit{\textbf{F}}_{\alpha^{\prime}\beta^{\prime}}}\hat{\psi}_{\beta}(\mathbf{r})\hat{\psi}_{\beta^{\prime}}(\mathbf{r})\\
&-3\hat{\psi}_{\alpha}^\dag(\mathbf{r})\hat{\psi}_{\alpha^{\prime}}^\dag(\mathbf{r})\left(\mathbf{\textit{\textbf{F}}}_{\alpha\beta}\cdot\mathbf{e}\right)\left(\mathbf{\textit{\textbf{F}}}_{\alpha^{\prime}\beta^{\prime}}\cdot\mathbf{e}\right)\hat{\psi}_{\beta}(\mathbf{r})\hat{\psi}_{\beta^{\prime}}(\mathbf{r})],
\end{split}
\end{equation}
where the strength $c_{dd}=\mu_{0}g_{F}^{2}\mu_{B}^{2}/(4\pi)$ with the magnetic permeability of the vacuum $\mu_0$, the Land\'e $g$ factor $g_F$ and  the Bohr magneton $\mu_B$, $\mathbf{e}=(\mathbf{r}-\mathbf{r^{\prime}})/|\mathbf{r}-\mathbf{r^{\prime}}|$ is the unit vector along $\mathbf{r}-\mathbf{r^{\prime}}$.

By adopting the standard mean-field approximation~\cite{KAWAGUCHI2012253}, GPEs was obtained as follow to solving the ground state of BEC~\cite{PhysRevA.93.053602, PhysRevA.102.013305},
\begin{equation}\label{eq:gpes}
\setlength\abovedisplayskip{3pt}
\setlength\belowdisplayskip{3pt}
i\hbar\frac{\partial\psi_{\alpha}(\mathbf{r})}{\partial t}= h_0\psi_{\alpha}+\textbf{B}_{e} \cdot \textbf{F}_{\alpha\beta}\psi_{\beta},
\end{equation}
where spin-independent term $h_0=T+V+c_{0}n$, $T=-\hbar^2\nabla^2/(2M)$ and $n=\sum_{\alpha}|\psi_{\alpha}|^2$ the total density. $\boldsymbol{B}_e(\boldsymbol{r})$ denotes an effective magnetic field origin from spin-dependent interactions,
\begin{equation}\label{eq:eff}
\textbf{B}_{e}({\textbf{r}})=c_{2}\textit{\textbf{f}}(\textbf{r})+c_{dd}\textbf{D}(\textbf{r}),
\end{equation}
where spin density $\textit{\textbf{f}}(\textbf{r})=\sum_{\alpha\beta}\psi_{\alpha}^\ast(\textit{\textbf{F}}_{\alpha\beta})\psi_{\beta}
$ and dipolar potential $\textbf{D}(\textbf{r})$'s $\mu=x,y,z$ component is,
\begin{equation}
\setlength\abovedisplayskip{3pt}
\setlength\belowdisplayskip{3pt}
D_{\mu}(\mathbf{r})=-\sum_{\nu=x,y,z}\int d\mathbf{r}^\prime\frac{(\delta_{\mu\nu}-3e_\mu e_\nu)\mathbf{\textit{\textbf{f}}}_{\nu}(\mathbf{r}^\prime)}{|\mathbf{r}-\mathbf{r}^\prime|^3}.
\end{equation}

Ground state and its dynamical behavior are obatined though out real and image time evolution of GPEs, respectively. 


The energy of the system can be divide into five parts,
\begin{equation}
	E = T + E_0 + E_2 +E_{dd} +E_v.
\end{equation}
In this study, the following notations are employed: $T$ denotes the kinetic energy; $E_0$ represents the corresponding $c_0$ term; $E_2$ signifies the spin exchange $c_2$ term; $E_{dd}$ indicates the energy of MDDI; and $E_v$ is the energy of the external field $V(r)$.

To simplify calculations, we utilize $a_{r}=\sqrt{\hbar/(M\omega_{0})}$ for length unit, $E_0=\hbar\omega_{0}$ for energy unit, gravity of the earth $g$ for acceleration unit, and $n_r=N/a_r^3$ for density unit.

\section{Off-axis spin vortex}
\label{sec:osv}
Spin vortex can be induced by its intrinc dipole-dipole interaction between atoms which has been investigated before 
~\cite{sadler2006spontaneous, PhysRevLett.100.170403, PhysRevLett.97.130404, PhysRevA.81.063623, PhysRevA.82.043627}.  In this paper the spin vortex we consider about is PCV~\cite{PhysRevLett.96.065302, KAWAGUCHI2012253, PhysRevA.100.033603, PhysRevResearch.3.013154, PhysRevA.94.063615}. 

We will briefly review the generation of PCV by MDDI. The MDDI energy, defined as the mean value of $\hat{H}_{dd}$, can be rewritten using the Fourier form of spin density and the real space position vector $\boldsymbol{r}$ as follows
\begin{equation}
	E_{dd}= \frac{1}{4\pi^2}c_{dd}\int d \textbf{\textit{k}}|\hat{\textbf{\textit{k}}}\cdot\tilde{\textbf{\textit{f}}}(\textbf{\textit{k}})|^{2}  -\frac{2\pi}{3}{c_{dd}}\int d \textbf{\textit{r}}|{\textbf{\textit{f}}}(\textbf{\textit{r}})|^2,
	\label{eq:edd}
\end{equation}
where $\boldsymbol{f}(\boldsymbol{k})$ is the Fourier transform of the spin density $\boldsymbol{f}(\boldsymbol{r})$.

As outlined in the paper~\cite{PhysRevA.105.063324}, the two terms of the right side of Eq.~\ref{eq:edd} were defined as $E'_{dd}$ and $E''_{dd}$ respectively. It was demonstrated that $E''_d$ and $E_2$ share the same form, with $E_2$ being defined as $E_2\equiv (c_2/2)\int d\boldsymbol{r}|f(\boldsymbol{r})|^2$. This results in a new definition of the spin exchange energy $E_2' = E''_d + E_2$.

It is evident that, in a traditional PCV, $E''_d = 0$ due to the  axisymmetry around the z-axis of the system. It is thus demonstrated that the existence of non-axisymmetry external potential, i.e., microgravity, results in the breaking of the spin formulation($\hat{\textbf{\textit{k}}}\cdot\tilde{\textbf{\textit{f}}}(\textbf{\textit{k}})=0$)~\cite{ueda2010fundamentals}. In this section, a toy model is employed to evaluate the effect of microgravity.

In this study, we consider a special case, $U_0$ and $\sigma$ are sufficiently large to ensure the entire BEC is a cyclic with the radius $\sigma$, one-dimensional case. The energy difference between the SMA and PCV states with microgravity is obtained to specify the effect of microgravity. We emphasize that the density of this toroidal BEC will not be uniform due to the effects of microgravity. The calculation of density is achieved through the Thomas-Fermi(TF) approximation, with the spin interactions and kinetic energy terms being ignored~\cite{pethick2008bose}. 

When $a=0$, the total density $n$ is found to be uniform, $n=N/(2\pi\sigma)$.  It is further assumed that the external potential is $V(R)$, and the chemical potential is denoted by $\mu$. In the TF approximation, the uniform density have the formulation,
\begin{equation}
	\bar n(\vec{R})=\frac{\mu-V(R)}{c_0},
\end{equation}
where $\vec{R}=\sigma \vec \varphi$ is the position vector with $\varphi=|\vec{\varphi}|$ is the azimuthal angle. In this situation, we consider a microgravity $a$, the density is
\begin{equation}
	\begin{aligned}
	n(\vec{R})&=\frac{\mu-V(R)-Max}{c_0}\\
	&=\bar{n}-bx\\
	&=\bar{n}-b\sigma\cos\varphi,
	\end{aligned}
	\label{eq:n}
\end{equation}
where $b = Ma/c_0$, the density-density interaction is represented by $c_0$. We emphasize that the chemical potential remains the same in the uniform case due to the conservation of the number of atoms, $N = \int d\vec{R} n(\vec{R})$.  Eq.~\ref{eq:n} is only applicable when $b\sigma < \bar{n}$.

\begin{figure}
	\centering
	\includegraphics[width=3.25in]{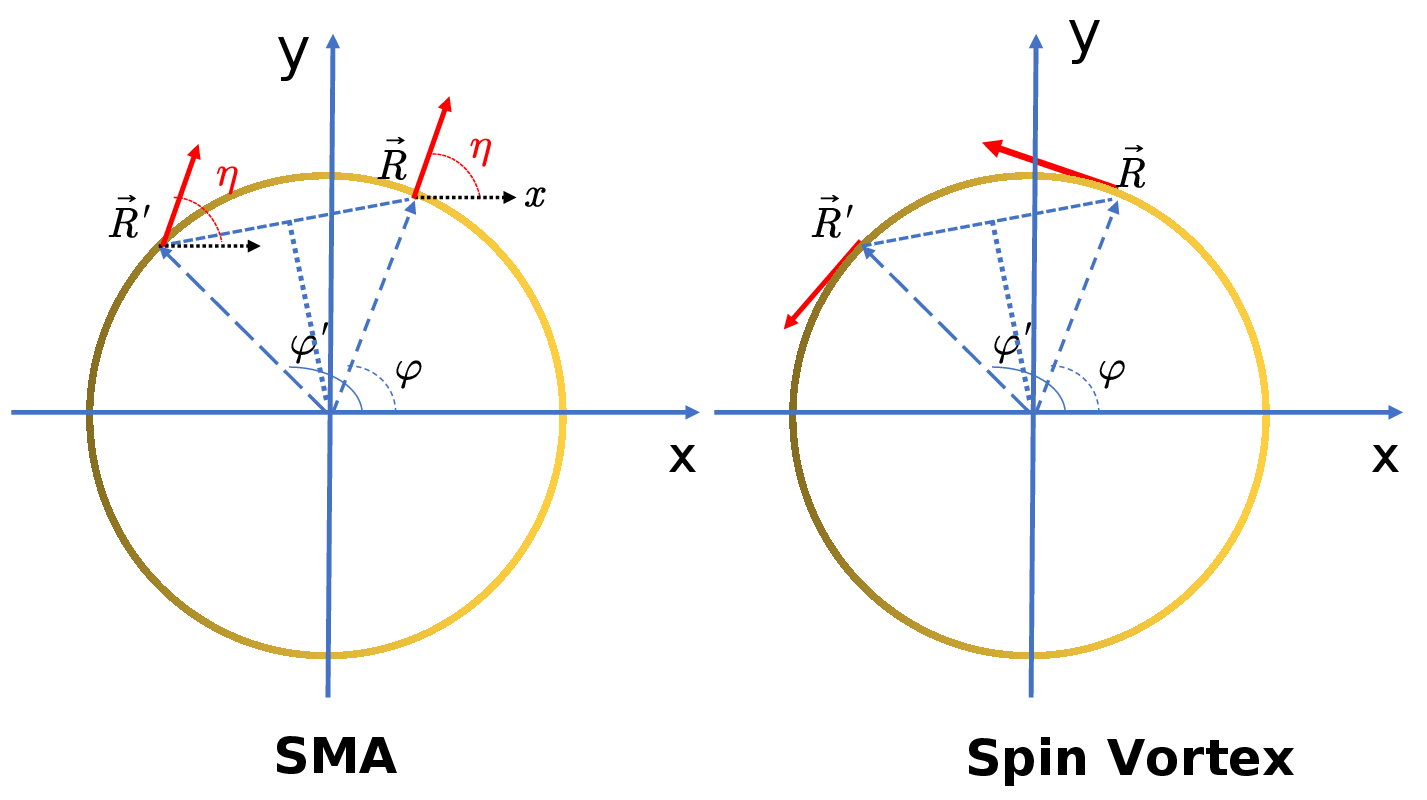}
	\caption{The SMA and spin vortex in a one-dimensional ring with the radius $\sigma$. The red vector represents the spin density at two positions. The intensity of the background color indicates density. Due to microgravity along the x-axis, the density is higher on the left than on the right. }
	\label{fig:od}
\end{figure}

Rewrite the Eq.~\ref{eq:n} as follow,
\begin{equation}
	n(\varphi)=\frac{N}{2\pi \sigma}(1-b'\cos\varphi ),
\end{equation}
where $b'=\frac{2\pi \sigma^2b}{N}\propto a$. Given the specified density, the energy of the two cases illustrated in Fig.~\ref{fig:od}, can be determined by the phases of three components. Because the minimum energy of $E_2'$ can be achieved in SMA and spin vortex configurations, respectively, by adjusting the density ratio to $n_{\pm 1}(\varphi) = n_0(\varphi)/2=1/4n(\varphi)$, and the phase $\phi_m(\varphi),(m=-1,0,1)$ to the ralation~\cite{PhysRevA.60.1463},
\begin{equation}
	\phi_{1}+\phi_{-1}-2\phi_0=0.
	\label{eq:phase}
\end{equation}
For a typical axisymmetry PCV the phase satisfies $\phi_m=-m\varphi +\varphi_m$, and $\phi_m=\varphi_m$ is constant for SMA ~\cite{PhysRevA.60.1463,PhysRevLett.97.130404,PhysRevLett.97.020401,PhysRevA.66.011601,PhysRevLett.94.160401,zhang2003mean,doi:10.1143/JPSJ.70.1604}.  However, for a non-axisymmetric spin vortex, the $\phi_m$ may not be linear with respect to the azimuthal angle $\varphi$. We therefore use the general form 
\begin{equation}
\phi_m=g_m(\varphi)+\varphi_m,
\label{eq:phi_m}
\end{equation}
with $g_m(\varphi)$ is the function of $\varphi$, under the constrain Eq.~\ref{eq:phase}.

As demonstrated in Fig.~\ref{fig:od}, the density and spin density of SMA and spin vortex are presented by background color of ring and red vectors. $\eta$ represents the angle between spin desnity and x axis in SMA. The intensity of the background color indicates the density. We refer to non-axisymmety spin vortex as off-axis spin vortex(OSV), just like the off-axis vortex of the scalar BEC~\cite{jezek2008metastable, yuce2010off, yuce2011critical, PhysRevA.70.063620, pethick2008bose}.

Since the density distribution and the external field remain the same for two states, and $E_2$ has reached the minimum, we only need compare the $T$ and $E_{dd}$ between two states drawn in Fig.~\ref{fig:od}. It is easy to calculate the $T$ and $E_{dd}$ of the SMA with microgravity $a$, because the density and the phase are obtained. $\phi_m=\varphi_m$ is a constant for SMA, implying that $T^S$ is only dependent on the density distribution, and $E_{dd}^S$ is straightforward to obtain. The superscripts $S$ and $V$ represent the SMA and spin vortex, respectively.

Due to the one dimension ring feature, the superfluid velocity $\boldsymbol{v}_{\pm1}$ of $\vert \pm 1\rangle $ are~\cite{KAWAGUCHI2012253}
\begin{equation}
	\boldsymbol{v}_{\pm1}=\frac{\hbar }{M}\boldsymbol{\nabla}\phi_{\pm1},
	\label{eq:v}
\end{equation}
where its radial component $v_{\rho}=0$, and its azimuthal represented by $v_{\varphi}$. In the ground state, transition between the spin components, due to spin exchange, reaches dynamic equilibrium, while the velocity field is non-zero in the $+1$ or $-1$ component, satisfying the respective continuity equation,
\begin{equation}
	\frac{\partial n_{\pm1}}{\partial t}+\nabla\cdot[n_{\pm1}\boldsymbol{v}_{\pm1}]=0.
	\label{eq:continue}
\end{equation}
We assume that $\boldsymbol{v}_{\pm1}(\varphi)=\frac{\hbar}{M\sigma}w_{\pm1}(\varphi)\hat{e}_{\varphi}$, where $w_{\pm1}=g_{\pm1}'$. Using Eq.~\ref{eq:v}, we find,
\begin{equation}
	\frac{\partial \phi_{\pm1}}{\partial \varphi}=w_{\pm1}(\varphi).
	\label{eq:phi}
\end{equation}
Substituting Eq.~\ref{eq:phi} into Eq.~\ref{eq:continue} yields the solution,
\begin{equation}
	w_{\pm1}(\varphi)=\frac{\mp1\sqrt{1-b'^2}}{1-b'\cos\varphi}.
	\label{eq:w}
\end{equation}
Another trivial solution of Eq.~\ref{eq:continue} is $w_{\pm} = constant$, which is a typical SMA state. As $a$ approaches zero, $w_{\pm1}$ reduces to $\mp1$, which is consistent with an axisymmetric spin vortex~\cite{PhysRevA.105.063324}.

We expand Eq.~\ref{eq:w} by $b'$ when $b'\ll1$,
\begin{equation}
	w_{\pm1}=\mp(1+b'\cos\varphi),
\end{equation} 
and then find the solution of Eq.~\ref{eq:phi} is,
\begin{equation}
	\phi_{\pm1}(\varphi)=\mp(\varphi-b'\sin\varphi)+\varphi_{\pm1},
	\label{eq:phase in OSV}
\end{equation}
with $\phi_0=\varphi_0$, $\varphi_m$ is a constant and $\varphi_1+\varphi_{-1}-2\varphi_0=0$.

Now that the density of SMA and OSV are known, the direction of the spin density, $\boldsymbol{f}(\varphi)$, depends on the phase shift, $\phi_0(\varphi)-\phi_{1}(\varphi)$~\cite{PhysRevLett.97.020401}, which can be adjusted by $\varphi_m$ under the constraint, Eq.~\ref{eq:phase}.  For an OSV, the spin density can be calculated using the phase and density distributions; however, the direction of the SMA spin density is unknown. Due to the non-uniform density distribution, the $E_{dd}$ of the SMA depends on the parameter $\eta$ in Fig. 1. We then minimize the $E_{dd}^S$ by varying the value of the parameter $\eta$ to determine the spin density of the SMA with microgravity $a$. We find that, when  $\eta=0$, $E_{dd}^S$ reaches its minimum, meaning that $\boldsymbol{f}(\varphi)$ is aligned with the direction of microgravity. The details are summarized in Appendix~\ref{apd:ring}.

Once the density and spin density have been obtained, it is easy to find the energy difference $\Delta E $,
\begin{equation}
	\Delta E = -\frac{c_{dd}N}{8\pi\sigma^3}I_{dd}+\frac{c_{dd}Nb'^2}{128\pi \sigma^3}A_{dd}+\frac{\hbar^2\sqrt{1-b'^2}}{4M\sigma^2},
	\label{eq:energy shift}
\end{equation}
with
\begin{equation}
	\begin{aligned}
	I_{dd}&=\int_{\varphi_i}^{\varphi_f}d\varphi_{-}\frac{3-2\sin^2\varphi_-}{4\vert \sin\varphi_-\vert^3},\\
	A_{dd}&=\int_{\varphi_i}^{\varphi_f}d\varphi_{-}\frac{8-6\cos(2\varphi_-)-\cos(4\varphi_-)}{\vert \sin\varphi_-\vert^3},
	\end{aligned}
\end{equation}
where $I_{dd},A_{dd}>0$ are depend on the trunction length $r_c$~\cite{PhysRevA.63.053607}. An assumption is made that $r_c$ is the mean distance between two atoms in the ring, i.e. $r_c=2\pi \sigma/N$, and its coresponding trunction angle $\varphi_c=2\pi/N$ with $\varphi_i=\varphi_c, \varphi_f=2\pi-\varphi_c$.

The Eq.~\ref{eq:energy shift} indicates that the smaller the $a$($b'\propto a$) and $\sigma$, the more favorable the conditions are for the generation of OSV ground state.  Since we have assumed that this is a one-dimensional ring, this condition requires that both $\sigma$ and $U_0$ are large enough. Furthermore, to generate a spin vortex, $\sigma$ must be greater than the dipole healing length, which is given by $\xi_{dd}=\hbar/\sqrt{2Mc_{dd}\bar{n}}$~\cite{PhysRevLett.97.130404,PhysRevA.82.043627,PhysRevA.105.063324}.  To obtain a OSV ground state with microgvarity $a$, $\Delta	E<0$ shoud be satisfy, we get the range of $\sigma $ from Eq.~\ref{eq:energy shift},
\begin{equation}
	\sigma<\sigma_c\equiv4M\left(\frac{c_{dd}NI_{dd}}{8\pi}-\frac{c_{dd}Nb'^2A_{dd}}{128\pi}\right)/(\hbar\sqrt{1-b'^2}),
\end{equation}
and $\sigma>\xi_{dd}$ is another condition. However, once $\sigma_c<\xi_{dd}$ or $\sigma_c<0$, i.e., $\frac{c_{dd}I_{dd}}{8\pi}-\frac{c_{dd}Nb'^2A_{dd}}{128\pi}<0$, the generation of OSV is impossible. Those constrain may be satisfy when $N$ large enough and $a$ small enough~\cite{PhysRevA.105.063324}.  

As a result of the analysis above, it can be concluded that there may be a range of parameters, i.e., $a$ and $\sigma$ where the OSV can exist as a ground state. And  the phase diagram spanned by $a$ and $\sigma$ can be obtained by more accurate numerical calculations in next section. Here, we assume that the phase diagram has been obtained. We will now continue to use the toy model to find the rough behaviour of the phase diagram. The relationship between $\sigma$ and $a$ of a contour $\Delta E=constant$ is dependent on microgravity, as easily determined from Eq.~\ref{eq:energy shift}
\begin{equation}
	\frac{d\sigma}{db'}=\frac{b'\sigma[2Q_1\sqrt{1-b^{\prime2}}-Q_2\sigma]}{\sqrt{1-b^{\prime2}}[3Q_3+3Q_1b^{\prime2}+2Q_2\sigma\sqrt{1-b^{\prime2}}]},
	\label{eq: sigma with gravity}
\end{equation}
where,
\begin{equation}
	\begin{aligned}
		Q_1=  \frac{c_{dd}N}{128\pi}A_{dd}; Q_2=  \frac{\hbar^2}{4M};Q_3=-\frac{c_{dd}N}{8\pi}I_{dd}.
	\end{aligned}
\end{equation}
When $b'\rightarrow0$ and $\sigma$ large enough, i.e., $2Q_1\sigma\sqrt{1-b'^2}\gg 3Q_3+3Q_1b'^2$ in Eq.~\ref{eq: sigma with gravity}, the Eq.~\ref{eq: sigma with gravity} is simplify as,
\begin{equation}
	\frac{d\sigma}{db'}=\frac{-b'(2Q_1-Q_2\sigma)}{2Q_2}.
	\label{eq: sigma with gravity appro}
\end{equation}
The solution of Eq.~\ref{eq: sigma with gravity appro} is 
\begin{equation}
	\sigma=\sigma_m(1-Ae^{\frac{-b^{\prime2}}{4}}),
	\label{eq:sigma and a}
\end{equation}
where $\sigma_m=2Q_1/Q_2$, $A=CQ_2/2Q_1$ and $C$ is a parameter depend on inital condition $\Delta E=constant$. 

In this section, the one-dimensional toy model is employed to determine the condition of generation a OSV ground state by calculating the energy difference between SMA and OSV. When N is large enough, existence parameters $a$ and $\sigma$ ensure the condition $\xi_{dd}< \sigma < \sigma_c$. This, in turn, causes the emergence of an OSV ground state. The relationship between $a$ and $\sigma$ has been determined for the contour $\Delta E=constant$.



\section{Numerical results on phase transition}
\label{sec:num}

\begin{figure}
\centering
\includegraphics[width=3.25in]{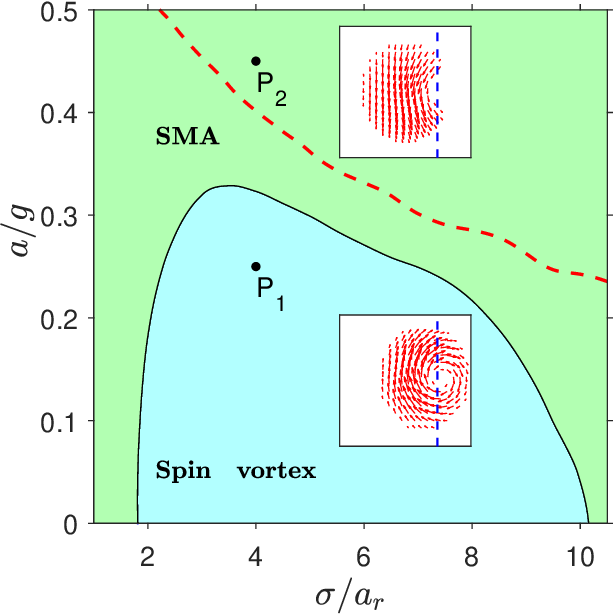}
\caption{The phase diagram of the dipolar $^{87}$Rb BEC ground state is presented here, spanned by microgravity acceleration $a$ and optical plug width $\sigma$ with its intensity $U_0 = 120E_0$. The solid black line in the color map denotes the phase transition line, the green region above the line corresponds to the SMA state, and the cyan region below is the spin vortex state. The red dashed line indicates the parameter selection at this BEC ``opening''. Which means the optical plug is not surrounded by BEC. The two small insets sketch the spin density of the two states in the $z=0$ plane, with the red arrows denoting the spin density and the length representing the amplitude. The blue dashed line corresponds to the position $x=0$, where the optical plug is added. The inset above corresponds to the spin density of the ground state in the $P_2$ parameter, where the optical plug is outside the BEC. The insert below illustrates the spin density distribution of the spin vortex state.Despite the fact that the spin is no longer aligned with the axisymmetric distribution, its configuration remains in a circular pattern centered on the core of the optical plug.}
\label{fig:pd}
\end{figure}

We found the ground state of the $^{87}$Rb dipolar spin-1 condensate by numerical solving coupled GPEs Eq.~(\ref{eq:gpes}). The FFT algorithm is employing to calculate the kinetic and dipolar terms~\cite{bao2003numerical,shi2018variational}. Due to the long-range MDDI nature, the truncation approach has been used to improve the numerical precession~\cite{PhysRevA.105.063324,PhysRevA.66.023613,PhysRevA.74.013623}.

We consider a high pancake dipolar BEC condition, whose trap frequency $\omega_{0}=2\pi\times 100$ Hz and trap aspect ratio $\lambda=20$. And the optical plug intensity $U_0=120\hbar\omega_{0}$. The phase diagram is spanned by two parameters, optical plug width $\sigma$ and microgravity $a$. The results are summaryed in Fig.~\ref{fig:pd}. 

\begin{figure}
	\centering
	\includegraphics[width=3.25in]{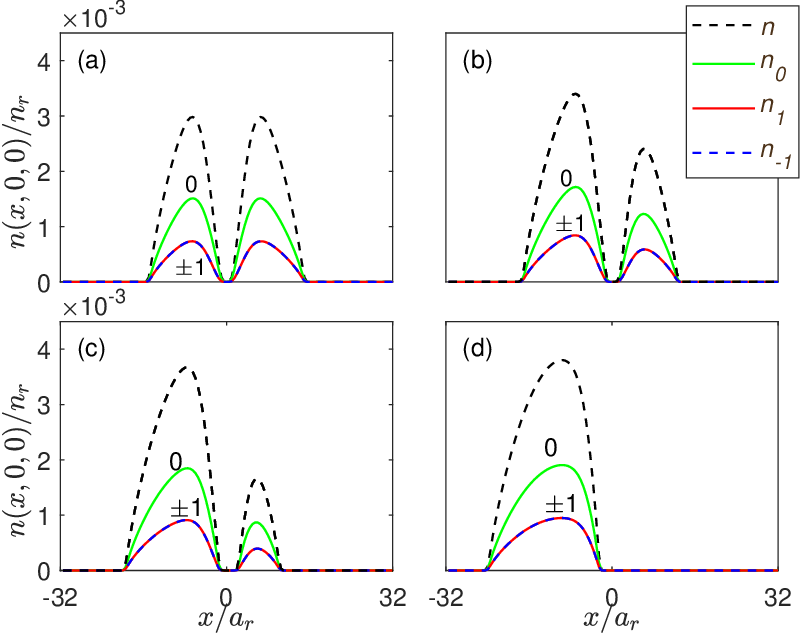}
	\caption{Density distribution of the components of the ground state on the $z=0,y=0$ line for each gravity with $\sigma=9a_r$. Plots (a)-(f) correspond to the ground state under parameters $a=0,0.1,0.2,0.4g$, respectively, and the other parameters are the same as in Fig.~\ref{fig:pd}. }
	\label{fig:density}
\end{figure}

In Fig.~\ref{fig:pd}, the green and cyan represent the single mode state (SMA) and the spin vortex state, respectively. The black solid line between the two colors indicates the phase boundary. In the SMA state, the direction of the spin density is uniform, and its amplitude is proportional to the density. The ground state at the $P_1$ point in the parameter space of Fig.~ref{fig:pd} has been selected to illustrate the spin distribution of a typical SMA state shown in upper inset in Fig.~\ref{fig:pd}, under microgravity and optical plug conditions.
The red vector of insets represent the spin density, and its length represents the amplitude. We chose the gorund state of $P_2$ point in Fig.~\ref{fig:pd} to exhibit the spin texture of the spin vortex state. The spin density forms a flux-closure structure, exhibiting a topological non-trivial spin density which can be found in Fig.~\ref{fig:od}. The blue dash line represent the position of optical plug. In fact, the presence of the optical plug at point $P_2$ ``outside'' the BEC implies the impossibility of the formation of a spin vortex around the plug.  As can be seen in the Fig.~\ref{fig:pd}, in the ground state of point $P_2$, the optical plug is indeed ``outside'' of the BEC. However, we require a more precise mathematical definition of what is meant by ``outside''. Here, we define the ``outside'' situation where the maximum density of the right of plug is less than $1\%$ of the total maximum density. The red dash line indicates the boundary between the ``outside'' and ``inside'' situations. The SMA state below the red dash line represent the case the optical plug inside the BEC, but not support the spin vortex ground state. In accordance with the qualitative analysis in Sec.~\ref{sec:osv}, the picture illustrates that $\sigma$ must be located between $\sigma_c$ and $\xi_{dd}$ to generate the spin vortex ground state. As microgravity $a$ increases, the range of $\sigma$ in the spin vortex state decreases, indicating that increasing microgravity $a$ does not promote the formation of the spin vortex ground state. The phase boundary near $a\rightarrow 0, \sigma\approx10a_r$ reflects Eq.~\ref{eq:sigma and a} in Sec.~\ref{sec:osv}.

We present the density distribution in $x$ direction ($y,z=0$) for four cases $a=0,0.1g,0.2g,0.4g$ from (a) to (d) in Fig.~\ref{fig:density} for a better understanding of the effect of micgravity. The width of optical plug $\sigma=9a_r$ in Fig.~\ref{fig:pd}. The anisotropic property of the MDDI is shown by $f_z(\mathbf r) =n_1-n_{-1}= 0$ for all cases, whereby spins align in the $x$-$y$ plane, which is the easy plane, for $\lambda \gg 1$~\cite{PhysRevLett.97.020401, PhysRevA.82.043627, PhysRevA.73.023602}. From Fig.~\ref{fig:pd}, it is straightforward to determine that (a) and (b) reflect the density distribution of the spin vortex state, whereas (c) and (d) indicate the SMA state. A typical axisymmetry polar-core vortex (PCV) is shown in (a)~\cite{PhysRevLett.97.020401}.It is clear that in each instance, the core region has a very low total density because of the optical plug. Moreover, the $n_0/n_1=2$ requirement holds true over the whole spatial domain. This result implies that $E_2$ achieves its minimum under such a potential, due to the polarization of spin,
\begin{equation}
\boldsymbol{\tilde f}(\boldsymbol{r})\equiv\frac{\boldsymbol{f}(\boldsymbol{r})}{n(\boldsymbol{r})},
\end{equation}
is maximally polarized, i.e., $|\boldsymbol{\tilde{f}}|=1$, this holds for all space. (b) A typical non-axial symmetry spin vortex state is indicated here. This state has been named off-axis spin vortex(OSV), in order to differentiate it from PCV. The presence of microgravity results in a lower density on the right side(positive x direction) of the optical plug in comparison to the left, thereby disrupting the axial symmetry surrounding the z-axis. It has been demonstrated that when microgravity increases to 0.2, the ground state undergoes a transition to SMA. In addition, the density distribution indicates that the spin density is fully polarized. In comparison with (a) and (b), the density on the right side of the optical plug is lower.

\begin{figure}[!t]
	\centering
	\includegraphics[width=3.25in]{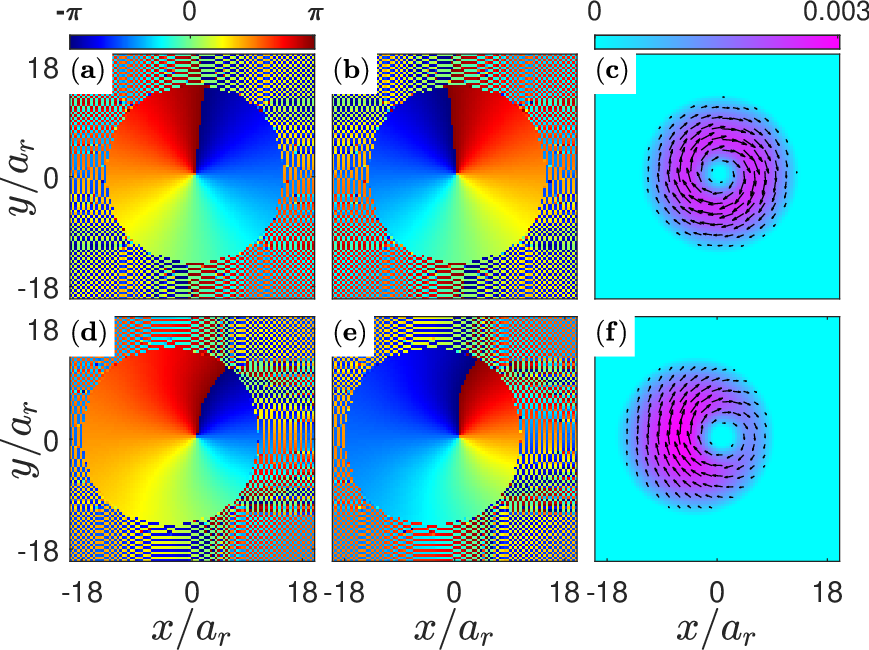}
	\caption{Off-axis spin vortex compared to axisymmetric spin vortex. The upper row (a)-(c) corresponds to the distribution of $\phi_{1}$, $\phi_{-1}$ and spin density distribution of the ground state under the parameter $a=0$, $\sigma=4a_r$, and the other parameters are the same as in Fig. ~\ref{fig:density}, where the ground state is an axisymmetric spin vortex state. (c) shows that the spin density arrangement in the $z=0$ plane has an axisymmetric closed circulation. The following rows (d)-(f) are the ground state at $a=0.2g$, and the other parameters are the same as those in the row above. The ground state at this parameter is an off-axis spin vortex state, and the phase $\phi_{\pm1}$ still exhibits topologically non-trivial behavior in space with a optical plug center, except that the vortex nucleus is now not geometrically centered in the region of the BEC, as can be seen in figure (f). The background colors in (e) and (f) are the total density distribution of the ground state at the respective parameters.}
	\label{fig:phase}
\end{figure}

As illustrated in  Fig.~\ref{fig:phase}, the phase and spin density of two types of spin vortex are presented. The top row of the figure illustrates the phase and spin density of axisymmetric spin vortex in the absence of microgravity. (a) and (b) represent the phase of the $\vert 1\rangle$ and $\vert -1\rangle$ component, respectively. The topological non-trival structure is clearly apparent in the $\vert 1\rangle$ and $\vert -1\rangle$ phases. Additionally, $\theta_{\pm1}$ is linearly related to the azimuthal angle. (a) and (b) exhibit winding numbers of $1$ and $-1$, respectively. The winding number is the phase change in space, after rotating around the optical plug in counterclockwise, by an integer multiple of $2\pi$. The $\vert\pm1\rangle$ phase of an OSV with $a=0.2g$ is shown in (d) and (e). The axisymmetry is absent, but the topological structure remains around the optical plug. Furthermore, the presence of microgravity causes the phase $\theta_{\pm1}$ to not be linear with the azimuthal angle, which is consistent with the qualitative analysis in Sec.~\ref{sec:osv}. The phase of $\vert 0\rangle$ is constant and is not displayed here. The (c) and (f) display the spin density in the $z=0$ plane. Because the direction of spin density depends on $\Delta\varphi\equiv \theta_0-\theta_{1}$~\cite{PhysRevA.105.063324}, the topological non-trivial phase distribution leads to the topological non-trivial spin density distribution. The background color of (c) and (f) respresent the total density distribution of two states. Microgravity causes the core of the spin vortex to shift away from the center of the BEC, as shown in (f).

\begin{figure}
	\centering
	\includegraphics[width=3.25in]{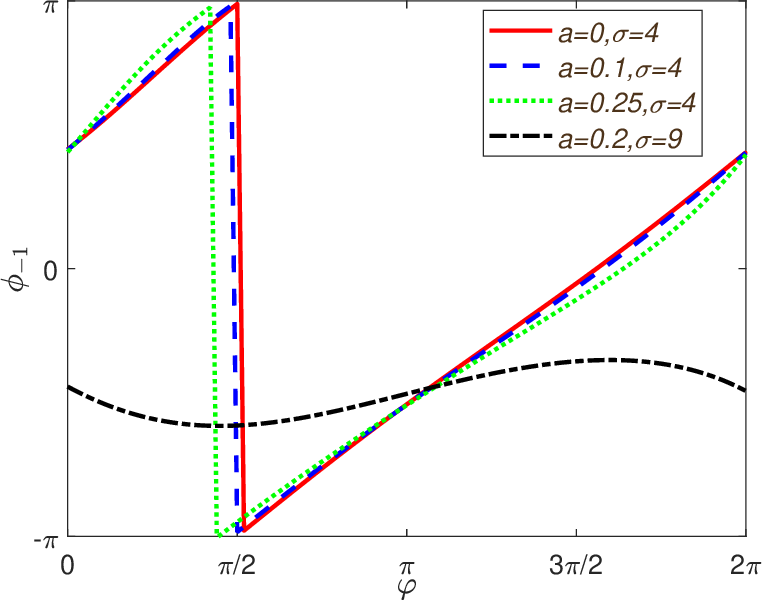}
	\caption{
		The phase of the $-1$ component is studied as a function of spatial azimuth. The path corresponding to the radius where the density of the right of optical plug in each ground state is largest is selected. For an optical plug width $\sigma=4a_r$, the phase of the $-1$ component of the ground state corresponds to the red solid line, blue double line, and cyan dashed line in the figure under the parameters of $a/g=0,0.1,0.25$. Those states are spin vortex ground state.  The black dotted line depicts the ground state of $\sigma=9$ and $a=0.2g$ which is a typical SMA state. The remaining parameters coincide with those shown in Fig.~\ref{fig:phase}. }
	\label{fig:Ut}
\end{figure}

As previously observed, $\theta_{\pm1}$ does not appear to be linear with the azimuthal angle $\varphi$. We show the phase of $\vert-1\rangle$ along the trajectory of the cycle around the center of the optical plug. The radius of the cycle is the distance between the point of maximum density on the left side of the optical plug and its center. The results are summary in Fig.~\ref{fig:Ut}. 

The red solid line depicts the phase distribution of the axisymmetric spin vortex state, which is the ground state for $a=0$ and $\sigma=4a_r$. The other parameters are the same as in FIg.~\ref{fig:phase}. Except for the phase jump point(from $\pi$ to $-\pi$), $\theta_{-1}$ has a linear relationship with the azimuth angle $\varphi$. The $\theta_{1}$ is positive linear dependent on $\varphi$, and the jumping point is 1, hence the winding number is $+1$. The blue dashed line and green dotted line show the ground states for $a=0.1g,\sigma=4a_r$ and $a=0.25g,\sigma=4a_r$, respectively. These are OSV states, whose phase distributions are roughly the same to those of the axisymmetric case but do not vary linearly with azimuth angle. By comparing the solid, dash, and dot lines, increasing $a$ resulted in a larger nonlinear effect, which is consistent with the qualitative analysis in Sec.~\ref{sec:osv}. The dot dash black line represents a typical SMA scenario in which $\theta_{-1}$ approximates a constant and the winding number is zero. 

\begin{figure}
	\centering
	\includegraphics[width=3.25in]{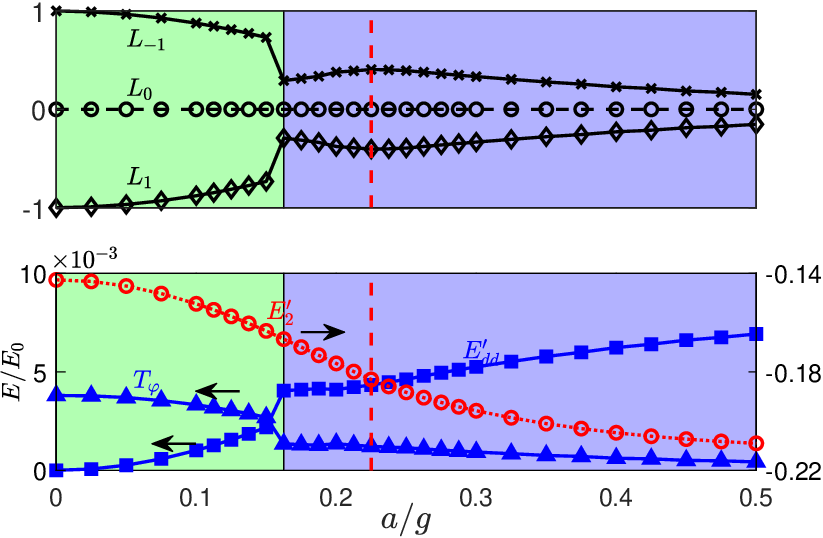}
	\caption{Diagram of the phase transition due to microgravity. As gravity $a$ increases, the system change from the FCLSV structure in the spin-vortex state to the off-axis one and finally to the SMA state. The cyan region corresponds to the spin-vortex state, the blue region corresponds to the SMA, and the vertical red dashed line corresponds to the ``opening", implying that above this gravity, there are parts of the optical plug that are outside the BEC. At this point, the spin-vortex state of a circle around the optical plug does not exist. The top picture shows the average angular momentum $L_m$ of the atoms that are in each component, and the bottom figure shows $E_2‘$ (red dotted circle, coordinates correspond to the right), $E_{dd}’$ (blue squares) $T_{\varphi}$ (triangles) in the ground state. The parameters are the same as in Fig.~\ref{fig:Ut}, except that gravity is used as the horizontal coordinate parameter, $\sigma=9$.}
	\label{fig:a}
\end{figure}

For an axisymmetric spin vortex, the winding number of each spin component is proportional to its angular momentum~\cite{PhysRevLett.96.065302}, 
\begin{equation}
	L_{m}=\frac{\int d\textbf{r}\psi_{m}^{\ast}(\textbf{r})(-i\frac{\partial}{\partial\varphi})\psi_{m}(\textbf{r})}{\int d\textbf{r}|\psi_{m}(\textbf{r})|^2},\; (m=1, 0,-1)
\end{equation}
while the non-axisymmetry of BEC breaking this relationship due to the microgravity. Fortunately, $L_m$ also shows an abrupt change in phase boundary where we set the $\sigma=9a_r$, as shown in Fig.~\ref{fig:a}, caused by the change in winding number of $\theta_{\pm1}$. The critical point of the phase transition is $a_c=0.1625g$ in Fig.~\ref{fig:a}. The critical point $a_c$ separates the cyan and blue colors, which stand for the spin vortex and SMA, respectively. 
Phase transitions are caused by competition between the MDDI energy $E_{dd}$ and the kinetic energy along the azimuthal angle,
 $$T_{\varphi}=\sum_{m=-1,0,1} \int d\mathbf r \; \psi_m^*(\mathbf r)  \left(-\frac{\partial^2}{2\rho^2\partial \varphi^2}\right) \psi_m(\mathbf r),$$
as discussed in Sec.~\ref{sec:osv}. The fact that $\theta_{\pm1}$ is not linearly dependent on azimuthal angle implies that $\Delta\varphi$ is as well, which can be found in Fig.~\ref{fig:phase} and also confirmed in Sec.~\ref{sec:osv}. This indicates that the spin density is not entirely perpendicular to the radial direction towards the center of the optical plug. And results in a finite value of $E_{dd}'$ and a non-integer $L_{\pm1}$ in spin vortex state, which are zero and $\mp1$ in the axisymmetry spin vortex state ($a=0$), respectively. This behavior can be found in the lower of Fig.~\ref{fig:a}. Although the finite value $E_{dd}'$ in two states, the abrupt occur during the phase transition. Similar to $E_{dd}'$, the abrupt occur in $T_{\varphi}$ of Fig.~\ref{fig:a}, indicating a phase transition. The red dashed line represents a critical point $a_o=0.225g$ where the optical plug is located on the boundary of BEC. The area $a>a_o$ represents the ''outside'' situation, while the ''inside'' is on its left.  Obviously, there exists a range $a_c<a<a_o$ where the optical plug is surrounded by the BEC and the ground state is SMA. It demonstrates that improving microgravity does not promote the formation of spin vortex ground states as confirmed in Sec.~\ref{sec:osv}.

\section{Conclusions}
\label{sec:conc}

A disc dipolar spinor $^{87}$Rb BEC can benefit from the optical plug's assistance to generate the spin vortex's ground state. We use a one-dimensional model to qualitatively analyze how microgravity affects spin vortex ground state generation. We conclude that microgravity-induced phase transitions do occur. The spin vortex state, also known as OSV, is non-axisymmetric and occurs in microgravity. Unlike previous research, this OSVof this study is a stable ground state rather than a dynamical intermediate state~\cite{PhysRevA.93.053602}. Numerical solutions to the Gross-Pitaevskii equation allow for more precise calculations. We set the optical plug intensity $U_0=120E_0$ and obtain a phase diagram by adjusting the microgravity $a$ and the optical plug width $\sigma$. This phase diagram shows two states, SMA and the spin vortex state. Under microgravity, the spin vortex state have an off-axis structure. Our study explains how microgravity reacts to phase transitions and proposes a method for creating spin vortex in a microgravity environment. This also implies that microgravity may be measured using quantum phase transitions.


\begin{acknowledgments}
This work is supported by the National Natural Science Foundation of China under Grant No. 12274331. 
\end{acknowledgments}

\appendix
\section{Derivation of Eq.~(\eqref{eq:energy shift})}
\label{apd:ring}

As mentioned in Sec.~\ref{sec:osv}, we have already obtain the density of two case in Fig.~\ref{fig:od}: $n(\varphi)=\frac{N}{2\pi \sigma}(1-b'\cos\varphi)$ with $b'=\frac{2\pi \sigma^2Ma}{Nc_0}\propto a$. The wavefuncition is expressed as $\psi_m=1/\sqrt{2\pi}\sqrt{1-b'\cos\varphi}\exp(i\theta_m)$ $(m=0,\pm1)$ with the relation $\phi_1+\phi_{-1}-2\phi_0=0$. Thus the spin density is,
\begin{equation}
	\boldsymbol{f}(\varphi)=n(\varphi)(\cos\Delta\varphi,\sin\Delta\varphi),
\end{equation}
where $\Delta\varphi=\theta_0-\theta_{1}$. 

The energy difference, denoted by the symbol $\Delta E$, is calculated between two local energy minima, the SMA and the OSV, which are denoted by the superscripts $S$ and $V$, respectively. Because the spind density is maximium polarization $|\boldsymbol{f}(\varphi)/n(\varphi)|=1$ is satified in two cases, the energy difference of contact interaction is $E_{2}^V+E_{0}^V-(E_{2}^S+E_{0}^S)=0$. The potential energy difference is also zero. What we need to calculate are kinetic and dipolar energy. 

For the SMA in Fig.~\ref{fig:od}, $\Delta\varphi^S=\eta$ is a constant. The MDDI energy of SMA is,
\begin{equation}
	\begin{aligned}
		E_{dd}^S=\frac{c_{dd}N}{128\pi \sigma^3}\int_{\varphi_i}^{\varphi_f}d\varphi_{-} 
\frac{-4+3b^{\prime2}\cos(2\eta)-2b^{\prime2}\cos(2\varphi_-)}{|\sin\varphi_-|^3}
	\end{aligned}
\end{equation}
where $\varphi_-=(\varphi-\varphi')/2$, $\varphi$ and $\varphi'$ are the azimuthal angle of two atoms.  And $\varphi_i=\varphi_c$, $\varphi_f=2\pi-\varphi_c$ with $\varphi_c=2\pi/N$ is the trunction angle~\cite{PhysRevA.63.053607}. Clearly, $E_{dd}^S$ reaches its minimum when the spin density is aligned with the direction of microgravity, i.e., when $\eta=0$. So, we can set $\theta_m=0$ $(m=\pm1,0)$ for this SMA state.

 In the case of the OSV, when $b'=0$ the $\Delta\varphi$ should be return to a PCV, i.e., $\Delta\varphi=\pm\pi/2+\varphi$,$+$ and $-$ in front of the $\pi/2$ represent  counterclockwise and clockwise spin texture of the spin vortex, respectively~\cite{PhysRevA.105.063324}.  Here, we chose $\Delta\varphi=\pi/2+\varphi$ to match the spin vortex in Fig.~\ref{fig:od}. As the result, we obtain that.
 \begin{equation}
 	\theta_{\pm1}=\pm(\varphi-b'\sin\varphi),
 \end{equation}
 with $\theta_m=0$. And $\Delta\varphi$ of the OSV can be derive from the Eq.~\ref{eq:phase in OSV} , 
\begin{equation}
\Delta\varphi^V = \frac{\pi}{2}-b'\sin\varphi+\varphi.
\label{eq:Deltaphi V}
\end{equation}
Utilizing the equation before, the spin density $\boldsymbol{f}(\varphi)$ can be obtained, the dipolar energy of the OSV is,
\begin{equation}
	\begin{aligned}
		E_{dd}^V=&-\frac{c_{dd}N}{64\sigma^3\pi^2}\int d\varphi \int d\varphi'\frac{(1-b'\cos\varphi)(1-b'\cos\varphi')}{|\sin\frac{\varphi-\varphi'}{2}|^3}\\
		&\times(3\cos [b'(\sin\varphi+\sin\varphi')]\\
		&+\cos(\varphi'-\varphi+b'\sin\varphi-b'\sin\varphi')).
	\end{aligned}
	\label{eq:MDDI PCV with gravity}
\end{equation}

The kinetic energy is,
\begin{equation}
	\setlength\abovedisplayskip{3pt}
	T=\sum_{m}\int \sigma d\varphi \psi_{m}^{\ast}\left(-\frac{1}{2\sigma^2} \frac{\partial^2}{\partial\varphi^2}\right) \psi_{m} .
\end{equation}
As the density and the phase are known for two states, the kinetic energy difference is,
\begin{equation}
	\Delta T=\frac{\hbar^2N\sqrt{1-b^{\prime2}}}{4M\sigma^2}.
	\label{eq:kin in ring with gravity}
\end{equation}
The MDDI energy difference is,
\begin{equation}
\begin{aligned}
\Delta E_{dd}=&E_{dd}^V-E_{dd}^S\\
=&-\frac{c_{dd}N}{8\pi \sigma^3}\int_{\varphi_i}^{\varphi_f} d\varphi_{-}\frac{3-2\sin^2(\varphi_{-})}{4|\sin(\varphi_{-})|^3}\\
&-\frac{c_{dd}Nb^{\prime2}}{\pi \sigma^3}\int_{\varphi_i}^{\varphi_f} d\varphi_{-}\frac{-8+6\cos(2\varphi_{-})+\cos(4\varphi_-)}{128|\sin\varphi_-|^3}\\
=&-\frac{c_{dd}N}{8\pi \sigma^3}I_{dd}+\frac{c_{dd}Nb^{\prime2}}{128\pi\sigma^3}A_{dd},
\end{aligned}
\label{eq:MDDI shift with gravity}
\end{equation}
where $A_{dd}$ and $I_{dd}$ are determined by the trunction angle,
\begin{equation}
	\begin{aligned}
		I_{dd}&=\int_{\varphi_i}^{\varphi_f}d\varphi_{-}\frac{3-2\sin^2\varphi_-}{4\vert \sin\varphi_-\vert^3},\\
		A_{dd}&=\int_{\varphi_i}^{\varphi_f}d\varphi_{-}\frac{8-6\cos(2\varphi_-)-\cos(4\varphi_-)}{\vert \sin\varphi_-\vert^3},
	\end{aligned}
\end{equation}
with $\varphi_i=\varphi_c$, $\varphi_f=2\pi-\varphi_c$, $\varphi_{\pm}=(\varphi\pm\varphi^{\prime})/2$. And $\varphi_c$ is trunction angle. We set $\varphi_c=2\pi/N$ which represent the average angle between two atoms in the ring.
As the result, the total energy differerce is,
\begin{equation}
	\Delta E = -\frac{c_{dd}N}{8\pi\sigma^3}I_{dd}+\frac{c_{dd}Nb'^2}{128\pi \sigma^3}A_{dd}+\frac{\hbar^2\sqrt{1-b'^2}}{4M\sigma^2}.
\end{equation}


\end{document}